\documentstyle[12pt,aaspp4]{article}

\received{13 April 1988}
\accepted{4 June 1988}
\slugcomment{ApJ Letters in press.}
\lefthead{Tomisaka}
\righthead{Collapse-Driven Outflow in Star-Forming Molecular Cores}

\begin{document}

\title{Collapse-Driven Outflow in Star-Forming Molecular Cores}
\author{Kohji Tomisaka}
\affil{Faculty of Education and Human Sciences,
 Niigata University, 8050 Ikarashi-2, 
 Niigata 950-2181, Japan}
\authoremail{tomisaka@ed.niigata-u.ac.jp}

\begin{abstract}
Dynamical collapses of magnetized molecular cloud cores are studied with 
 magnetohydrodynamical simulations from the run-away
 collapse phase to the accretion phase. 
In the run-away collapse phase, a disk threaded by magnetic field lines is 
 contracting due to its self-gravity and its evolution is well expressed by 
 a self-similar solution.
The central density increases greatly in a finite time scale and reaches
 a density at which an opaque core is formed at the center.
After that, matter accretes to the newly formed core (accretion phase).
In this stage,  a rotationally supported disk is formed
 in a cloud core without magnetic fields. 
In contrast, the disk continues to contract in the magnetized cloud core, 
 since the magnetic fields transfer angular momentum from the disk.  
Its rotation motion winds up the threading magnetic field lines.
Eventually, strong toroidal magnetic fields are formed and 
 begin to drive the outflow, even if there is no toroidal field
 component initially.
Bipolar molecular outflows observed in protostar candidates are naturally
 explained by this model. 
\end{abstract}


\section{Introduction}

Star forming interstellar cores often indicate molecular bipolar outflow
(for a review see Bachiller 1996).
A number of driving mechanisms are proposed:
 a rotating disk plus threading magnetic fields (disk-driven wind model:
 Uchida \& Shibata 1985; Pudritz \& Norman 1986; Ouyed \& Pudritz 1997) and   
 interaction of stellar and disk magnetospheres (boundary layer-driven wind
 model: Shu et al 1994).
As for the former model, when a disk with (sub-)Keplerian rotation speed is
 threaded by magnetic fields, it winds up the magnetic fields and generates
 strong toroidal component.
Resultant magnetic pressure gradient ejects matter from the disk.
In contrast, Shu and collaborators (1994) proposed the interaction
 between stellar magnetosphere and disk magnetic field.
Along the open fields where these two magnetospheres interact,
 disk matter is ejected by the centrifugal force.

Since rotation motion plays an important role in the dynamical contraction
 phase of the cloud core  and thus it generates strong toroidal
 magnetic fields, the matter ejection seems to be driven these two.
If the outflow occurs in a process of the dynamical contraction
 of the magnetized cloud core, it is necessary to investigate
 the process of gravitational contraction and outflow as a whole.
However, the previous works only considered steady states or short-range time
 evolutions of these phenomena.
Thus, in this Letter,
 we investigate the possibility of outflow driven by the self-gravitational
 contraction using `nested-grid' MHD simulations.
 
Analytic  (Larson 1969; Penston 1969; Hunter 1977; Whitworth \& Summers 1985)
 and numerical studies (Foster \& Chevalier 1993) have revealed
 that the dynamical collapse of a non-magnetic isothermal cloud
 is divided into two phases: the former ``run-away collapse'' phase
 and the latter ``accretion'' phase.
In the former phase, the cloud shows non-homologous contraction and the central
 density increases greatly in a finite time scale, while in the latter 
 accretion phase the infalling gas accretes onto a central high-density body
 formed in the run-away collapse phase. 
The latter phase corresponds to the ``inside-out collapse'' (Shu 1977) 
 which is realized when accretion begins at the center of 
 a hydrostatic cloud core.
For rotating (non-magnetic) clouds (Norman, Wilson, \& Barton 1980;
 Narita, Hayashi, \& Miyama 1984) 
 and magnetized (non-rotating) clouds (Scott \& Black 1980; for recent
 progress, see Tomisaka 1995), these two phases exist.

The centrifugal wind model by Blandford \& Payne (1982) predicts
 that cold disk matter could be accelerated by sufficiently strong
 magnetic fields
 when the angle between the field lines and the disk is smaller than
 $60^\circ$.
The structure of the cloud undergoing the run-away collapse is well
 described with a self-similar solution and the angle is predicted
 as $45^\circ$ with assuming that a  disk is thin and the magnetic field
 is current-free outside the disk (Basu 1997; Nakamura \& Hanawa 1997).
Therefore, if a magnetized cloud is rotating, outflow may naturally 
 be driven at some epoch in the self-gravitating contraction. 
To study the possibility, 
 we performed a two-dimensional ideal MHD 
 simulation assuming the cylindrical symmetry.
In this Letter, we report that the outflow appears to emanate from
 the disk in the latter ``accretion phase.''
  
\section{Model}

We begin our simulation from an infinitely long, cylindrical, rotating,
 isothermal cloud in hydrostatic balance, that is, the gravitational
 force is counterbalanced
 with the pressure force, the centrifugal force, and the Lorenz force.
Here we assume the rotation axis coincides with the cylinder ($z-$)axis.
As the initial state, only poloidal magnetic fields are taken into
 account and their strength is assumed proportional to the square root of 
 local gas density.
We add small density perturbations on it to initiate contraction.
The initial hydrostatic state is represented by two parameters
 which characterize the structure:
 $\alpha$ (the ratio of magnetic pressure to the thermal one
 for poloidal field component)  and
 $\Omega_0$ (the angular rotation speed at the center of the cloud).
Distributions of density, magnetic flux density, and rotation speed in
 the radial ($r-$)direction are taken identical with Matsumoto,
 Nakamura \& Hanawa (1994), although  we assume no toroidal magnetic
 field component  $B_\phi$ initially. 
The size of the numerical box is chosen to agree with the wavelength
 of the most unstable mode of the self-gravitational instability 
 $\lambda_{\rm MGR}$ (Matsumoto et al. 1994).
Quantities are scaled with the isothermal sound speed $c_s$, 
 the surface pressure (or surface density) $p_s=c_s^2 \rho_s$
 and the corresponding scale-height $H=c_s/(4\pi G \rho_s)^{1/2}$.
Typical values would be $c_s = 200 {\rm m~s^{-1}}$,
 $\rho_s = 100 {\rm cm^{-3}}$,
 $H = 0.36 {\rm pc}$, and thus the time-scale is normalized by
 the free-fall time as $\tau_{\rm ff}=(4\pi G \rho_s)^{-1/2}$ $\simeq 1.75
 {\rm Myr}(\rho_s / 100 {\rm cm^{-3}})^{-1/2} $.
Using these scalings, initial radial distributions of density, magnetic
 fields and angular rotation speed are written as 
$\rho(r)=(\rho_c/\rho_s)\{1+[\rho_c/\rho_s-2(\Omega_0 \tau_{\rm ff})^2]
                           /[8(1+\alpha)]\times (r^2/H^2)\}^{-2}\rho_s$,
$(B_z,B_r)=([2\alpha \rho(r)/\rho_s]^{1/2},0)\rho_s^{1/2}c_s$ and
$\Omega=\Omega_0 [\rho(r)/\rho_c]^{1/4}$.
As model parameters, $\alpha$, $\Omega_0$ and central density $\rho_c$ 
 are chosen as 
 $\alpha=0.5$, $\Omega_0=5/\tau_{\rm ff}$, and $\rho_c=100\rho_s$.

Since the non-homologous gravitational contraction necessitates progressively
 finer spatial resolution, ``nested grid technique'' (Burger \& Oliger 1984)
 is adopted,
 in which a number of grid systems with different spacings are prepared; 
 the center of the cloud is covered with a finer grid and the
 global structure is traced by a coarser grid simultaneously. 
We use 15 levels of grids from L0 (the coarsest) to L14 (the finest).
Grid spacing of L$n$ is chosen as a half of that of L$n-1$ and  
 each level of grids uses $64\times 64$ grid points.
Thus, the smallest grid (L14's) spacing is $\simeq 10^{-6}$ times
 smaller than  $\lambda_{\rm MGR}$.
We applied periodic boundary conditions on the upper and lower boundaries and the
 fixed one on the outer boundary.
Details of the numerical method have been described in separate papers 
 (Tomisaka 1996a,b).

\section{Results}

\subsection{Run-away Collapse Phase}

The evolution is similar to that of magnetized cloud with 
 no rotation (Tomisaka 1995, 1996a; Nakamura, Hanawa, \& Nakano 1995).
First, the cylindrical cloud breaks into prolate spheroidal fragments
 which is elongated along the cylinder axis ($z$-axis).
As long as the amplitude of density perturbation is small $\delta \rho
 / \rho \lesssim 1$, the shape keeps essentially identical with the most unstable 
 eigen-function of the gravitational instability.
Next, gas begins to fall mainly along the magnetic field, and forms a disk
 perpendicular to the magnetic fields.  
The disk formed in a run-away collapse is far from static balance.
However, the flow is controlled under the magnetic field and centrifugal
 force which lead to disk structure (pseudo-disk: Galli \& Shu 1993). 
At $t=0.5908 \tau_{\rm ff}$, 
 at that time the central density $\rho_c$ reaches $10^5\rho_s$,
 a shock wave is formed parallel to the disk (Figs.1a \& 1b).
It breaks into two waves at the epoch when the central density reaches 
 $\rho_c \sim 10^6 \rho_s$: an outer wave front propagates outwardly
 as a fast-mode MHD shock (seen near $z\sim 0.02 H$)
 and an inner one does inwardly as a slow-mode MHD shock
 (seen near $z\sim 7\times 10^{-3} H$) reaching the equator.
The rotation angular speed $\Omega$ and toroidal magnetic field component
 $B_\phi$ also jump crossing the wave fronts.
Fast rotation and thus strong $B_\phi$ are observed in a restricted region
 between these two fronts.
Since the disk contracts in the radial direction, it has a larger $\Omega$
 than outside the disk.
In this configuration, magnetic fields transfers the angular momentum 
 from the disk and redistributes it into the same magnetic flux 
 tube (magnetic braking).
However, from this numerical simulation it is shown that the angular
 momentum is confined into a region inside the fast-mode MHD shock fronts.

Although there exists a number of shock waves ($z\sim 0.02 H$, 
 $5\times 10^{-3}H$, and $2\times 10^{-4}H$) propagating in the $z$-direction,
 no such discontinuity is found in the $r$-direction (Figs.1c \& 1d).
However, a kind of modulation is seen around a power-law distribution as
 $\rho(z=0,r)\propto r^{-2}$, which is related to the formation
 of multiple shock waves in the $z$-direction
 (this is found by Norman et al (1980) for the case of rotating
 isothermal cloud).
The run-away collapse continues till $t\simeq 0.6 \tau_{\rm ff}$
 and the central density would increase greatly if the isothermal
 equation of state continues to be valid.
However, when the central density exceeds $10^{10}{\rm cm}^{-3}$,
 the central part of the core becomes optically thick for
 the thermal radiation from dusts and its temperature begins to rise 
 (for example, Masunaga, Miyama, \&  Inutsuka,1998).

\subsection{Accretion Phase}

In the accretion phase, gas continues to accretes onto the newly
 formed opaque core.
We mimic the situation by using a double polytrope composed of
 an isothermal one for low
 density as $p=c_s^2\rho$ for $\rho< \rho_{\rm crit}$
 and a harder one for high density as 
 $p=p(\rho_{\rm crit})(\rho/\rho_{\rm crit})^\Gamma$ for
 $\rho> \rho_{\rm crit}$.
We take $\rho_{\rm crit}=10^8 \rho_s=10^{10}{\rm cm}^{-3}(\rho_s/100{\rm cm}^{-3})$
 and $\Gamma=5/3$.
By virtue of this assumption we can follow further evolution.
Figure 2a shows the structure captured by L8 
 just after the isothermal equation of state is broken.
A disk, which runs vertically, contracts radially and gas motion
 drags and squeezes the magnetic field lines to the center.
The rotation speed $v_\phi$ at that time is not more than the radial infall
 speed $v_r$ which is in the range of $(2-3) \times c_s$ (see Fig. 1d).

In the accretion phase,
 the inflow speed is accelerated at least to $(4-5) \times c_s$
 and as a result magnetic field lines are strongly dragged and squeezed
 to the center (see Fig.1c of Tomisaka 1996b).
Beside these, the rotation speed is accelerated as the infall proceeds.
Just after the state shown in Figure 1 with the highest central concentration,
 $v_\phi$ reaches $\simeq 4c_s$ at the distance of $r\sim 3 \times 10^{-4} H$
 inside of which a nearly spherical core is formed supported mainly by thermal 
 pressure.   
The rotation velocity $v_\phi$ exceeds the radial velocity $v_r$
 in the accretion phase, while $v_\phi$ was smaller
 than $v_r$ (see Fig. 1d) in the run-away collapse phase.
From this epoch, outflow from the disk is observed.

Figure 2 shows the structures before (a) and after (b) the outflow begins.
The physical time scale between Figures 2a and 2b is approximately
 equal to $\simeq 1000 {\rm yr} (\rho_s/100 {\rm cm^{-3}})^{-1/2}$.
The directions of the magnetic fields in the disk is much affected
 by the radial inflow.
They are pushed and squeezed to the center.
It is shown that in the seeding region of the outflow
 the angle between a magnetic field line and the disk plane decreases
 from $60^\circ - 70^\circ$ (Fig.2a) to $10^\circ - 30^\circ$ (Fig.2b).
In Figure 2b, it is indicated that the outflow, which reaches
 $z_s \sim \pm 0.003 H \simeq \pm 200 {\rm AU} (H/0.36{\rm pc})$, 
 is confined into two expanding bubbles whose centers are 
 at $z_c \sim \pm 0.0015 H$.
In the bubble, the strength of toroidal component of magnetic fields $B_\phi$
 is larger than that of the poloidal one $(B_z^2+B_r^2)^{1/2}$.
Especially at the outer boundary of the bubble
  $|B_\phi|$ is $(4-5)$ times stronger than $(B_z^2+B_r^2)^{1/2}$ (see also the
 upper panel of Fig. 3).
Further, the magnetic pressure is much larger than the thermal one.
Outside of the bubble, inflow continues.  
The islands of strong toroidal magnetic fields which are formed just outside
 the disk push the poloidal fields and have an effect to strengthen the 
 hourglass structure in the poloidal field (Ouyed \& Pudritz 1997).

To see the seeding region of the outflow closely,
 we plot a close-up view captured by L10 in Figure 3.
(Seeding region, from which outflow is ejected, moves outwardly with time.
In the disk,
 the ratio of the centrifugal force to the gravity is no more than 
 $\simeq 0.45$.
It indicates this is far from centrifugal balanced state.)
This figure shows that the outflow is ejected from the disk near
 $r\simeq 5\times 10^{-4} H \simeq 40 {\rm AU} (H/0.36{\rm pc})$, which is
 5 times larger than the scale length for the critical density
 $\rho_{\rm cr}=10^{10}{\rm cm^{-3}}$.
Figure 3 clearly shows that (1) magnetic field lines are almost parallel
 to the disk surface near the disk.
(2) gas inflow in the disk is not disturbed by the outflow from
 the disk and continues to reach the central core which is supported by 
 the thermal pressure.
(3) outflow occurs along magnetic field lines
 whose inclination angle is in the range of $45^\circ$ - $60^\circ$.
This corresponds to the region where $|B_\phi|$ is dominant
 over the poloidal component (upper panel).

Consider the main outflow ejected from $r\simeq 5\times 10^{-4} H$
 in Figure 3.
Numerical simulations (Kudoh, Matsumoto, \& Shibata 1998;
 Ouyed \& Pudritz 1997) indicate that the ratio of Alfven speed
 to the rotation speed is controlling how the matter is accelerated.
Since this ratio is small as 0.3 -- 1 in the seeding region
 $r \sim  5\times 10^{-4} H$, it is natural that the toroidal magnetic 
 field is strongly  amplified just outside of the disk.
In this case, the magnetic pressure gradient is considered to
 play an important role to accelerate the gas after the Alfven 
 points located near $|z| \sim  10^{-4} H$.
(In contrast, the matter is centrifugally driven when the ratio is large.) 
However, even in the present model, gas seems to escape from the disk 
 with being driven centrifugally.  

\section{Discussion}

Inflow and outflow rates are measured at the outer boundary of L10 
 numerical box as the surface integral of the mass flux density as
 $\int \rho {\bf v}\cdot {\bf n} dS$.
The inflow rate is equal to
 $2.7\times 10^2 c_s^3/4\pi G \simeq 4.1\times 10^{-5} M_\odot{\rm yr}^{-1}
 (c_s/200{\rm m~s^{-1}})^3$, which is almost constant in the accretion phase.
It is worth noting that this is much larger than the standard accretion
 rate $12.25 c_s^3 / 4 \pi G$ predicted for the 
 singular isothermal sphere model (Shu 1977).
This  is due to the extra inflow velocity accelerated in the run-away 
 collapse phase (Tomisaka 1996b).
The outflow begins at $t\simeq 0.5998 \tau_{\rm ff}$
 and the outflow rate increases with time.
At the stage shown in Figure 3, it reaches  $83 c_s^3/4\pi G$
 $\simeq 1.2\times 10^{-5} M_\odot{\rm yr}^{-1}(c_s/200{\rm m~s^{-1}})^3$.
It is worth notice that the outflow rate 
 attains $\simeq 1/3$ of the inflow rate.
The ratio of outflowing mass to inflowing one decreases
 from L10 to L8.

Linear momentum outflow rate is also measured as the surface integral 
 of the momentum flux density $\int \rho v_z^2 dS$ over the upper and lower
 boundaries of L10.
This increases with time from the epoch of $t\simeq 0.5998 \tau_{\rm ff}$
 and reaches $\simeq 330 c_s^4/4\pi G$ 
 $\simeq 10^{-5} M_\odot{\rm km~s^{-1} yr^{-1}}
 (c_s/200{\rm m~s^{-1}})^4$ at the epoch shown in Figure 3.
This is not inconsistent with the momentum outflow rate
 observed around Class 0 low-mass YSOs 
 from $^{12}$CO $J=2-1$ line observation (Bontemps et al. 1996)  as
 $3\times 10^{-6}- 5\times 10^{-4} M_\odot{\rm km~s^{-1} yr^{-1}}$.
The maximum speed of outflow $(7-8)\times c_s$ is approximately
 equal to the Kepler speed of the seeding region, and it seems to be
 accelerated with increasing central mass and thus the Kepler speed.
Therefore, the momentum outflow rate is likely to increase further.

Why does the outflow occur only in the ``accretion phase''?
Rotation velocity becomes dominant over the radial velocity
 only in the accretion phase.
This is also indicated by a recent study using a self-similar solution for
 contracting isothermal rotating disk (Saigo \& Hanawa 1998;
 see their moderately rotating model of $\omega=0.3$).
Further, toroidal magnetic fields develop only in the accretion phase.

This seems to come from the fact that the run-away collapse is
 fast and as a result the  toroidal magnetic fields can not be developed
 by rotation.
In other words,
 from numerical simulations, the angular rotation speed near the center
 $\Omega_c(t)$ is proportional to the central free-fall rate as 
 $\Omega_c\simeq (0.2-0.4)\times (2\pi G \rho_c)^{1/2}$
(Matsumoto, Hanawa, \& Nakamura 1997).
Therefore, the angle that a gas element  
 rotates in this free-fall time is as large as 
 $\theta \sim \Omega_c /(2\pi G \rho_c)^{1/2} \sim 0.2-0.4$ radian.
Since the disk rotates only at $\sim 0.2-0.4$ radian,
 it is concluded that 
 the magnetic field can not be wound much in the run-away collapse phase.

In the accretion phase,
 we have found a rotating ring nearly in a hydrostatic balance
 for a model with no magnetic fields ($\alpha=0$).
Since the angular momentum in the disk is transferred to the
 outer region by the magnetic effect, a model with poloidal magnetic fields
 indicates a structure different from the non-magnetic disk.
Magnetized disk consists of an innermost nearly hydrostatic
 core and a contracting rotating disk.
  
Blandford and Payne (1982) have revealed that 
 a cold gas which rotates in a Kepler disk can escape from the stellar gravity
 by the effect of magnetic fields.
The condition to escape is that the angle between the magnetic field lines
 and the disk is smaller than 60$^\circ$,
 below which the magnetic field line can transfer enough angular momentum
 to the gas element to escape from the gravity of the central point mass.
Although this is valid for the point mass, the angle seems to play
 an important role for the outflow phenomenon.
Decrease in the angle is mainly due to the disk accretion (inflow)
 and drag. 
Therefore, it is concluded that
 the outflow is driven by a rotating contracting cloud core 
 only in the accretion phase.
 
\acknowledgments
I would like to thank Dr. F. Nakamura for careful reading the manuscript and
 Dr. R. Pudritz, referee, for his comments to improve the paper.  
Numerical simulations were performed by Fujitsu VPP300/16R supercomputer 
 at the Astronomical Data Analysis Center of the
 National Astronomical Observatory, Japan.

\onecolumn
\clearpage
\section*{Figure Captions}
\begin{figure}[h]
\caption{Distributions of the density 
 and infall speed  along the $z$-axis are plotted
 in (a) $\log \rho(z,r=0)$  and (b) $|v_z(z,r=0)|$  respectively.
Those along the equatorial plane are also shown in (c) $\log \rho(z=0,r)$
 and (d) $|v_r(z=0,r)|$.
Snapshots are made at $t=0.5233\tau_{\rm ff}$ ($\rho_c\sim 10^3\rho_s$ A),
$t=0.5725\tau_{\rm ff}$ ($\rho_c\sim 10^4\rho_s$ B),
$t=0.5908\tau_{\rm ff}$ ($\rho_c\sim 10^5\rho_s$ C),
$t=0.5977\tau_{\rm ff}$ ($\rho_c\sim 10^6\rho_s$ D),
$t=0.5994\tau_{\rm ff}$ ($\rho_c\sim 10^{8}\rho_s$ E),
$t=0.5996\tau_{\rm ff}$ ($\rho_c\sim 10^{8.8}\rho_s$ F), and
$t=0.5997\tau_{\rm ff}$ ($\rho_c\sim 10^{9.1}\rho_s$ G).
Rotation velocity distribution along the equatorial plane
 is plotted for the epoch G in (d).
}
\end{figure}
\begin{figure}[h]
\caption{Left: isodensity lines, magnetic field lines, and velocity vectors
 are plotted for the state when the central density reaches $10^{8.8} \rho_s$ 
 ($t=0.5996\tau_{\rm ff}$).
In contrast to the usual usage, the $z$-axis is placed horizontally and
 the $r$-axis is vertically.
Right: the same as (a) but for the state when the central density reaches
 $10^{10} \rho_s$  ($t=0.6002\tau_{\rm ff}$).
Both are Level 8.
Physical time passed between these two is equal to 1000 yr
 $(\rho_s/100{\rm cm^{-3}})^{-1/2}$.}
\end{figure}
\begin{figure}[h]
\caption{Close-up view of the seeding region of the outflow.
False colors represent 
 the ratio of the  toroidal magnetic pressure $B_\phi^2/8\pi$ to
 the poloidal one  $(B_z^2+B_r^2)/8\pi$ in the upper panel
 and the density distribution $\rho$ in the lower panel, respectively.
Velocity vectors ($v_z$, $v_r$) and magnetic field lines ($B_z$, $B_r$)
 are also plotted.
This figure shows Level 10 which has 4-times finer spatial resolutions than
 the previous one ($-10^{-3}H \le z \le 10^{-3}H$ and
 $0 \le r \le 2\times 10^{-3}H$).}
\end{figure}
\end{document}